\documentclass[runningheads]{llncs}
\usepackage{amssymb,amsmath}
\usepackage{pstricks}
\usepackage{pst-node}
\usepackage{epsfig}
\usepackage{latexsym}
\begin{document}
\pagestyle{headings}

\newcommand{\lang}{\ensuremath{\mathcal{L}}}
\def\paradef#1{\ensuremath{%
               \dr{}%
                  {{#1}^+\equiv\neg{#1}^-}%
                  {({#1}\equiv{#1}^+)\wedge(\neg{#1}\equiv{#1}^-)}}}
\def\paradefC#1#2#3{\ensuremath{%
                    \dr{}%
                       {({#1}\equiv{#1}_{#2}^+)\wedge(\neg{#1}\equiv{#1}_{#3}^-)}%
                       {({#1}\equiv{#1}_{#2}^+)\wedge(\neg{#1}\equiv{#1}_{#3}^-)}}}
\def\paradefPp#1#2{\ensuremath{%
                     \dr{}%
                       {({#1}\equiv{#1}_{#2}^+)}%
                       {({#1}\equiv{#1}_{#2}^+)}}}
\def\paradefPm#1#2{\ensuremath{%
                     \dr{}%
                       {(\neg{#1}\equiv{#1}_{#2}^-)}%
                       {(\neg{#1}\equiv{#1}_{#2}^-)}}}
\newcommand{\quantifier}{{\sf Q}}
\newcommand{\existence}{\mbox{\sc maxeq}}
\newcommand{\Rexistence}{\mbox{\sc Rdefext}}
\newcommand{\Cexistence}{\mbox{\sc Cdefext}}
\renewcommand{\existence}{\mbox{\sc defext}}
\newcommand{\choice}{\mbox{\sc choice}}
\newcommand{\Rchoice}{\mbox{\sc Rchoice}}
\newcommand{\Cchoice}{\mbox{\sc Cchoice}}
\newcommand{\skeptical}{\mbox{\sc skeptical}}
\newcommand{\Rskeptical}{\mbox{\sc Rskeptical}}
\newcommand{\Cskeptical}{\mbox{\sc Cskeptical}}
\newcommand{\theory}[1]{\mathit{Cn}({#1})}
\newcommand{\theoryl}[1]{\mathit{Cn}_\Sigma({#1})}
\newcommand{\Th}{\mathcal{T}}
\newcommand{\Mod}[1]{{\mathit{Mod}\/({#1})}}
\newcommand{\dr}[3]{\ensuremath{\frac{{#1}\,:\,{#2}}{#3}}}
\newcommand{\prereq}[1]{\ensuremath{\text{\it p}{\left( #1 \right)}}}
\newcommand{\justif}[1]{\ensuremath{\text{\it j}{\left( #1 \right)}}}
\newcommand{\conseq}[1]{\ensuremath{\text{\it c}{\left( #1 \right)}}}
\newcommand{\GD}[2]{\ensuremath{{G\!D}(#1,#2)}}
\newcommand{\GDI}[3]{\ensuremath{{G\!D}_{#3}}}
\newcommand{\GDIP}[3]{\ensuremath{{G\!D}'_{#3}}}
\newcommand{\CupGDI}[2]{\ensuremath{\bigcup_{i=0}^\infty\GDI{#1}{#2}{i}}}
\newcommand{\CupE}[1]{\ensuremath{\bigcup_{i=0}^\infty {#1}_i}}
\let\goodoldproof=\proof
\let\goodoldendproof=\endproof
\def\proof#1{\par\noindent\mbox{\bf Proof \ref{#1}} \ }
\def\endproof{\ \hfill\rule{2mm}{2mm}\medskip\par}
\newcommand{\Ext}[1]{\ensuremath{{\mathcal Ext}[{#1}]}}
\newcommand{\ExtH}[1]{\ensuremath{{\mathcal Ext}_h[{#1}]}}
\newcommand{\Cons}[1]{\ensuremath{{\mathcal Cons}[{#1}]}}
\renewcommand{\Cons}[1]{\ensuremath{{\mathcal Conseq}[{#1}]}}
\newcommand{\Consp}[1]{\ensuremath{{\mathcal Cons}_{G\leftarrow G'}[{#1}]}}
\renewcommand{\Consp}[1]{\ensuremath{{\mathcal Conseq}_{G\leftarrow G'}[{#1}]}}
\newcommand{\Cred}[1]{\ensuremath{{\mathcal Cred}[{#1}]}}
\newcommand{\Skept}[1]{\ensuremath{{\mathcal Skept}[{#1}]}}
\newcommand{\Prud}[1]{\ensuremath{{\mathcal Prud}[{#1}]}}
\newcommand{\CredH}[1]{\ensuremath{{\mathcal Cred_h}[{#1}]}}
\newcommand{\SkeptH}[1]{\ensuremath{{\mathcal Skept_h}[{#1}]}}
\newcommand{\PrudH}[1]{\ensuremath{{\mathcal Prud_h}[{#1}]}}
\def\paradefneu#1{\ensuremath{%
               \dr{}%
                  {{#1}^+\equiv\neg{#1}^-}%
                  {({#1}\equiv{#1}^+)\wedge(\neg{#1}\equiv{#1}^-)}}}
\def\paradefneuC#1#2#3{\ensuremath{%
                    \dr{}%
                       {({#1}\equiv{#1}_{#2}^+)\wedge(\neg{#1}\equiv{#1}_{#3}^-)}%
                       {({#1}\equiv{#1}_{#2}^+)\wedge(\neg{#1}\equiv{#1}_{#3}^-)}}}
\def\paradefneuPp#1#2{\ensuremath{%
                     \dr{}%
                       {({#1}\equiv{#1}_{#2}^+)}%
                       {({#1}\equiv{#1}_{#2}^+)}}}
\def\paradefneuPm#1#2{\ensuremath{%
                     \dr{}%
                       {(\neg{#1}\equiv{#1}_{#2}^-)}%
                       {(\neg{#1}\equiv{#1}_{#2}^-)}}}
\newcommand{\Fcons}[1]{\ensuremath{{\mathcal C}[{#1}]}}
\newcommand{\Fder}[1]{\ensuremath{{\mathcal B}[{#1}]}}
\newcommand{\Fmaxcons}[1]{\ensuremath{{\mathcal D}[{#1}]}}
\newcommand{\Fc}[1]{\ensuremath{{\mathcal D}_c[{#1}]}}
\newcommand{\Fp}[1]{\ensuremath{{\mathcal D}_p[{#1}]}}
\newcommand{\Fs}[1]{\ensuremath{{\mathcal D}_s[{#1}]}}
\newcommand{\Fcpm}[1]{\ensuremath{{\mathcal D}^{\pm}_c[{#1}]}}
\newcommand{\Fppm}[1]{\ensuremath{{\mathcal D}^{\pm}_p[{#1}]}}
\newcommand{\Fspm}[1]{\ensuremath{{\mathcal D}^{\pm}_s[{#1}]}}
\newcommand{\WEQ}{W^{\Eq}}
\newcommand{\Kdown}[2]{\mid #1 \mid_{#2}}
\newcommand{\Kup}[2]{\mid #1 \mid^{#2}}
\newcommand{\SAT}[0]{\mbox{\sc sat}}
\newcommand{\PSPACE}{{\rm PSPACE}}
\newcommand{\Sets}[1]{\ensuremath{S({#1})}}
\newcommand{\Eq}{\mathit{EQ}}
\newcommand{\Eqs}{\mathit{eq}}
\newcommand{\qPhi}[1]{\ensuremath{\Phi[{#1}]}}
\newcommand{\qPsi}[1]{\ensuremath{\Psi[{#1}]}}
\newcommand{\meq}[1]{\ensuremath{\Phi_{\mathit bm}[{#1}]}}
\renewcommand{\meq}[1]{\ensuremath{{\mathcal M}[{#1}]}}
\newcommand{\meqred}[1]{\ensuremath{{\mathcal M}^\ast[{#1}]}}
\newcommand{\Fexistence}[1]{\ensuremath{{\mathcal F}_{\mathit ext}[{#1}]}}
\newcommand{\Fchoice}[1]{\ensuremath{{\mathcal F}_{\mathit choice}[{#1}]}}
\newcommand{\Fskeptical}[1]{\ensuremath{{\mathcal F}_{\mathit skept}[{#1}]}}
\newcommand{\commadots}[0]{,\ldots ,}
\newcommand{\iec}[0]{i.e.,\ }
\newcommand{\Iec}[0]{I.e.,\ }
\newcommand{\ie}[0]{i.e.\ }
\newcommand{\verts}{{\,\vert \,}}
\newcommand{\co}{\mbox{\rm co-}}
\newcommand{\Pol}{{\rm P}}
\newcommand{\FPol}{{\rm FP}}
\newcommand{\NP}{{\rm NP}}
\newcommand{\FNP}{{\rm FNP}}
\newcommand{\np}{{\rm NP}}
\newcommand{\CONP}{\mbox{\rm co-}\NP }
\newcommand{\SigmaP}[1]{\Sigma_{#1}^{P}}
\newcommand{\PiP}[1]{{\Pi}_{#1}^{P}}
\newcommand{\SigmaE}[1]{\Sigma_{#1}^{exp}}
\newcommand{\PiE}[1]{{\Pi}_{#1}^{exp}}
\newcommand{\DeltaP}[1]{{\Delta}_{#1}^{P}}
\newcommand{\DeltaPpar}[1]{{\Delta}_{#1}^{P}}
\newcommand{\FDeltaP}[1]{{\rm F}{\Delta}_{#1}^{P} }
\newcommand{\PNP}{\Pol^\NP}
\newcommand{\PNPlog}{\PNP[O(\log n)]}
\newcommand{\FPNPlog}{\FPNP[O(\log n)]}
\newcommand{\PNPpar}{\Pol_\|^\NP}
\newcommand{\FPNPpar}{\FPol_\|^\NP}
\newcommand{\RPX}{{\rm RP}\cdot\FPNPpar}
\newcommand{\FPNP}{\FPol^\NP}
\newcommand{\PH}{{\rm PH}}
\newcommand{\ThetaP}[1]{{\Theta}_{#1}^{P} }
\newcommand{\DeltalogP}[1]{\DeltaP{#1}[O(\log n)] }
\newcommand{\FDeltalogP}[1]{\FDeltaP{#1}[O(\log n)] }
\newcommand{\parity}[1]{{\rm PARITY(#1)}}
\newcommand{\FNPOpt}{\FNP\mbox{\rm//OptP}[O(\log n)]}
\newcommand{\I}[0]{\ensuremath{\nu}}
\newcommand{\Ip}[0]{\ensuremath{\nu_2}}
\newcommand{\Iv}[1]{\I(#1)}
\newcommand{\Ivp}[1]{\Ip(#1)}
\newcommand{\Ivo}[0]{\I}
\newcommand{\val}[2]{\Ivo_{#1}({#2})}
\newcommand{\asgn}[1]{\ensuremath{I_{#1}}}
\renewcommand{\asgn}[1]{\ensuremath{\mathit{AS}[#1]}}
\newcommand{\MIN}[0]{\ensuremath{\mathit{min}}}
\newcommand{\MAX}[0]{\ensuremath{\mathit{max}}}
\newcommand{\sub}[2]{[#1 / #2]}
\newcommand{\subn}[4]{[#1 / #2 \commadots #3/#4]}
\newcommand{\subs}[3]{#1[#2 / #3]}
\renewcommand{\int}[1]{\ensuremath{\mathit{INT}[#1]}}
\newcommand{\var}[1]{\ensuremath{\mathit{var}(#1)}}
\newcommand{\QUIP}[0]{\mbox{\sf QUIP}}
\newcommand{\boole}{\texttt{boole}}
\newcommand{\egc}[0]{e.g.,\ }
\newcommand{\eg}[0]{e.g.,}

\sloppy
\title{Paraconsistent Reasoning via Quantified Boolean Formulas,
I: Axiomatising Signed Systems
\thanks{Originally published in proc. PCL 2002, a FLoC workshop;
eds. Hendrik Decker, Dina Goldin, J{\o}rgen Villadsen, Toshiharu Waragai
({\tt http://floc02.diku.dk/PCL/}).} 
\thanks{The work was partially supported by
   the Austrian Science Foundation under grant P15068.}
}
\author{%
 Philippe Besnard\inst{1}
 \and
 Torsten Schaub\inst{1} 
 \and
 Hans Tompits\inst{2}
 \and
 Stefan Woltran\inst{2}
}
\institute{%
Institut f\"ur Informatik,
 Universit\"at Potsdam,\\
 Postfach 90 03 27,
 D--14439 Potsdam,
 Germany\\
 \email{$\{$besnard$,$torsten$\}$@cs.uni-potsdam.de}
 \and
 Institut f\"ur Informationssysteme
184/3, 
 Technische Universit\"at Wien, \\
 Favoritenstra{\ss}e~9--11,
 A--1040 Vienna, Austria\\
 \email{$\{$tompits$,$stefan$\}$@kr.tuwien.ac.at}
 }

\maketitle
\begin{abstract}
~Signed systems were introduced as a general, syntax-independent framework for
paraconsistent reasoning, that is,\,
non-trivialised reasoning from inconsistent information.
In this paper, we show how the family of corresponding paraconsistent
consequence relations can be axiomatised by means of quantified Boolean
formulas.
This approach has several benefits.
First, it furnishes an axiomatic specification of paraconsistent reasoning
within the framework of signed systems.
Second, this axiomatisation allows us to identify upper bounds 
for the complexity of the different signed consequence relations.
We strengthen these upper bounds by providing strict complexity results
for the considered reasoning tasks. 
Finally, we obtain an implementation of different forms of paraconsistent reasoning by
appeal to the existing system \textsf{QUIP}.
\end{abstract}

\section{Introduction}
In view of today's rapidly growing amount and distribution of information,
it is inevitable to encounter inconsistent information.
This is why methods for reasoning from inconsistent data are becoming increasingly important.
Unfortunately, there is no consensus on which information should be derivable
in the presence of a contradiction.
Nonetheless, there is a broad class of consistency-based approaches that
reconstitute information from inconsistent data by appeal to the notion of
consistency.
Our overall goal is to provide a uniform basis for these approaches that makes
them more transparent and easier to compare.
To this end, we take advantage of the framework of quantified Boolean formulas (QBFs).
To be more precise, we concentrate here on axiomatising the class of so-called
\emph{signed systems}~\cite{Besnard98} for paraconsistent reasoning;
a second paper will deal with maximal-consistent sets and related approaches
(cf.~\cite{Cayrol98,Marquis02}).

Our general methodology offers several benefits:
First, we obtain uniform axiomatisations of rather different approaches.
Second, once such an axiomatisation is available, existing QBF solvers can be
used for implementation in a uniform setting.
The availability of efficient QBF solvers, like the systems described 
in~\cite{Cadoli98,Giunchiglia01,Feldmann00,Letz01}, makes such a rapid prototyping
approach practicably applicable.
Third, these axiomatisations provide a direct access to the complexity of
the original approach.
Finally, we 
remark that this approach allows us, in some sense, to
express paraconsistent reasoning in (higher order) classical propositional
logic and so to harness classical reasoning mechanisms from 
(a conservative extension of) propositional logic.

Our elaboration of paraconsistent reasoning is part of an encompassing
research program, analysing a large spectrum of reasoning mechanisms in
Artificial Intelligence, among them
nonmonotonic reasoning~\cite{Egly00c},
(nonmonotonic) modal logics~\cite{Eiter02},
logic programming~\cite{Egly00e,Pearce01},
abductive reasoning~\cite{Egly01b},
and
belief revision~\cite{desctowo01a}.

In order to keep our paper self-contained, we must carefully introduce the
respective techniques.
Given the current space limitations, we have thus decided to reduce the motivation and
rather concentrate on a thorough formal elaboration.
This brings us to the following outline:
Section~\ref{sec:background} lays down the formal foundations of our work,
introducing QBFs and Default Logic.
Section~\ref{sec:signed} is devoted to signed systems as introduced in~\cite{Besnard98}.
Apart from reviewing the basic framework, we provide new unifying
characterisations that pave the way for the respective encodings in QBFs,
which are the subject of Section~\ref{sec:red}.
This section comprises thus our major contribution:
a family of basic QBF axiomatisations that can be assembled in different ways
in order to accommodate the variety of paraconsistent inference relations
within the framework of signed systems.
We further elaborate upon these axiomatisations in
Section~\ref{sec:complexity} for analysing the complexity of the respective
reasoning tasks. 
Finally, our axiomatisations are also of great practical value since they allow for
a direct implementation in terms of existing QBF-solvers.
Such an implementation is described in Section~\ref{sec:discussion}, 
by appeal to the system \QUIP~\cite{Egly00c,Egly00e,Egly01b}. 

\section{Foundations}\label{sec:background}
\subsection{Preliminary Notation}

We deal with propositional languages and use the logical symbols
$\top$, $\bot$, $\neg$, $\vee$, $\wedge$, $\to$, and $\equiv$
to construct formulas in the standard way.
We write $\lang_\Sigma$ to denote a language over an alphabet
$\Sigma$ of \emph{propositional variables} or \emph{atoms}.
Formulas are denoted by Greek lower-case letters (possibly with subscripts).
Finite sets $T=\{\phi_1\commadots\phi_n\}$ of formulas 
are usually identified with the conjunction 
$\bigwedge_{i=1}^n\phi_i$ of its elements. 
The set of all atoms occurring in a formula $\phi$ is denoted by $\var{\phi}$.
Similarly, for a set $S$ of formulas, $\var{S}=\bigcup_{\phi\in S}\var{\phi}$.
The derivability operator, $\vdash$, is defined in the usual way.
The \emph{deductive closure} of a set $S\subseteq\mathcal{L}_\Sigma$ of
formulas is given by $\theoryl{S}=\{\phi\in\mathcal{L}_\Sigma\mid S\vdash\phi\}$.
We say that $S$ is \emph{deductively closed} iff $S=\theoryl{S}$.
Furthermore, $S$ is \emph{consistent} iff $\bot\notin\theoryl{S}$.
If the language is clear from the context, we usually drop the index
``$\Sigma$'' from $\theoryl{\cdot}$ and simply write $\theory{\cdot}$
for the deductive closure operator.

For formulas $\varphi$, $\phi$, and  $\psi$, we define \emph{positive} and 
\emph{negative occurrences} as follows:
\begin{itemize}
\item
  the occurrence of $\varphi$ in $\varphi$ is positive,
\item
  if $\varphi$ 
  occurs positively (negatively) in $\phi$, then
  the corresponding occurrence of $\varphi$ in 
  $\neg\phi$ and $\phi\to\psi$ is negative (positive),
\item
  if $\varphi$ 
  occurs positively (negatively) in $\phi$, then
  the corresponding occurrence of $\varphi$ in 
  $\phi\vee\psi$, $\phi\wedge\psi$, and $\psi\to\phi$ is positive (negative).
\end{itemize}
Given an alphabet $\Sigma$, we define a disjoint alphabet
$\Sigma^\pm$ as 
\(
\Sigma^\pm=\{p^+,p^-\mid p\in \Sigma\}
\).
For $\alpha \in \lang_\Sigma$, we define
\(
\alpha^\pm
\)
as the formula obtained from $\alpha$ by replacing each negative occurrence
of
\(
p
\)
by
\(
\neg p^-
\)
and by replacing each positive occurrence of
\(
p
\)
by
\(
p^+
\),
for each propositional variable $p$ in $\Sigma$.
For example
\(
(p  \wedge (p       \to q  ))^\pm
=
 p^+\wedge (\neg p^-\to q^+)
\).
This is defined analogously for sets of formulas.
Observe that for any set $T\subseteq\lang_\Sigma$, $T^\pm$ is consistent, 
even if $T$ is inconsistent.

\subsection{{Quantified Boolean Formulas}}\label{sec:logical}

Quantified Boolean formulas (QBFs) generalise ordinary propositional
formulas by the admission of quantifications over propositional
variables (QBFs are denoted by Greek upper-case letters).
Informally, a QBF of form $\forall p \, \exists q \, \Phi$
means that for all truth assignments of $p$ there is a truth
assignment of $q$ such that $\Phi$ is true.
For instance, it is easily seen that the QBF
$
\exists p\, \exists q \, ( (p \to q)
\wedge \forall r (r \to q))
$
evaluates to true.

The precise semantical meaning of QBFs is defined as follows.
First, some ancillary notation.
An occurrence of a propositional variable $p$ in a QBF $\Phi$ is \emph{free}
iff it does not appear in the scope of a quantifier
$\quantifier p$ ($\quantifier\in\{\forall,\exists\}$),
otherwise the occurrence of $p$ is \emph{bound}.
If $\Phi$ contains no free variable occurrences, then $\Phi$ is
\emph{closed}, otherwise $\Phi$ is \emph{open}.
Furthermore, we write $\Phi \subn{p_1}{\phi_1}{p_n}{\phi_n}$ to denote the
result of uniformly substituting each free occurrence of a variable $p_i$
in $\Phi$ by a formula $\phi_i$, for $1\leq i\leq n$.

By an  \emph{interpretation}, $M$, we understand a set of atoms.
Informally, an atom $p$ is true under $M$ iff $p\in M$.
In general, the truth value, $\val{M}{\Phi}$, of a QBF $\Phi$ 
under an interpretation $M$  is recursively defined as follows:
\begin{enumerate}
\item if $\Phi=\top$, then $\val{M}{\Phi} = 1$;
  
\item if $\Phi=p$ is an atom, then $\val{M}{\Phi} = 1$ if
  $p\in M$, and $\val{M}{\Phi} = 0$ otherwise;
  
\item if $\Phi=\neg \Psi$, then
  $\val{M}{\Phi} = 1 - \val{M}{\Psi}$;
  
\item if $\Phi=(\Phi_1\wedge \Phi_2)$, then
  $\val{M}{\Phi} = \MIN(\{\val{M}{\Phi_1},\val{M}{\Phi_2}\})$;
  
\item if $\Phi= \forall p \, \Psi$, then
  $\val{M}{\Phi} = \val{M} {\Psi \sub{p}{\top}
    \wedge \Psi \sub{p}{\bot}}$;
  
\item if $\Phi= \exists  p \, \Psi$, then
  $\val{M}{\Phi} = \val{M} {\Psi \sub{p}{\top} \vee
    \Psi \sub{p}{\bot}}$.
\end{enumerate}
The truth conditions for $\bot$, $\vee$, $\to$, and $\equiv$ follow
from the above in the usual way.
We say that $\Phi$ is \emph{true under $M$} iff $\val{M}{\Phi}=1$, 
otherwise $\Phi$ is \emph{false under $M$}. 
If $\val{M}{\Phi}=1$, then $M$ is a \emph{model} of $\Phi$. 
If $\Phi$ has some model, then $\Phi$ is said to be \emph{satisfiable}.
If $\Phi$ is true under any interpretation, then $\Phi$ is \emph{valid}.
As usual, we write $\models\Phi$ to express that $\Phi$ is valid.
Observe that a closed QBF is either valid or unsatisfiable, because
closed QBFs are either true under each interpretation or false under
each interpretation.
Hence, for closed QBFs, there is no need to refer to particular interpretations.
Two sets of QBFs (or ordinary formulas) are \emph{logically equivalent}
iff they possess the same models.

In the sequel, we use the following abbreviations in the context of QBFs: 
For a set $P = \{ p_1,\ldots,p_n\}$ of propositional 
variables and a quantifier $\quantifier\in\{\forall,\exists\}$, we 
let $\quantifier P \, \Phi$ stand for the formula 
$\quantifier p_1 \quantifier p_2\cdots\quantifier p_n\,\Phi$.
Furthermore,
for 
indexed sets 
$S=\{\phi_1,\ldots,\phi_n\}$ and $T=\{\psi_1,\ldots,\psi_n\}$ 
of formulas, $S\leq T$ abbreviates 
$\bigwedge_{i=1}^n (\phi_i \rightarrow \psi_i)$.

The operator $\leq$ is a fundamental tool for expressing certain tests 
on sets of formulas in terms of QBFs. 
In particular, we use $\leq$ for expressing the following task:
\begin{quote}
{Given finite sets $S$ and $T$ of formulas, compute all subsets 
$R\subseteq S$ such that $T\cup R$ is consistent.}
\end{quote}
This problem can be encoded by a QBF in the following way:
\begin{proposition}
\label{prop:sat-check}
Let $S = \{ \phi_1 \commadots \phi_n \}$ and $T$ be finite sets of formulas,
let $P=\var{S\cup T}$, and 
let $G= \{ g_1 \commadots g_n \}$ be a set of new variables. 
Furthermore,
for any $S'\subseteq S$, define the interpretation $M_{S'}\subseteq G$
such that $\phi_i\in S'$ iff $g_i\in M_{S'}$, for $1\leq i\leq n$.

Then, $T\cup {S'}$ is consistent iff the QBF
\[
\Fcons{T,S} = \exists P (T\wedge (G \leq S))
\]
is true under~$M_{S'}$.
\end{proposition}
\begin{theorem}\label{thm:trans:1}
Given the prerequisites of Proposition~\ref{prop:sat-check}, 
we have that ${S'}$ is a maximal subset of $S$ consistent with $T$
iff 
$M_{S'}$ is a model of the QBF
$$
\Fcons{T,S} \wedge \bigwedge_{i = 1}^n \Big( \neg g_i \to \neg\, 
\Fcons{T\cup\{\phi_i\},S\setminus\{\phi_i\}}\Big).
$$
\end{theorem}

\subsection{{Default Logic}}\label{sec:default}

The primary technical means for dealing with ``signed theories'' is
\emph{default logic} \cite{Reiter80},
whose central concepts are \emph{default rules} along with their induced
\emph{extensions} of an initial set of premises. 
A default rule (or \emph{default} for short)
\[
\dr{\alpha}{\beta}{\gamma}
\]
has two types of antecedents: a \emph{prerequisite} $\alpha$ which is
established if $\alpha$ is derivable
and a \emph{justification} $\beta$ which is established if $\beta$ is
consistent.
If both conditions hold, the \emph{consequent} $\gamma$ is concluded by
default.
For convenience, we denote the prerequisite of a default $\delta$ by 
$\prereq{\delta}$, its justification 
by
$\justif{\delta}$, and its consequent by $\conseq{\delta}$.
Accordingly, for a set of defaults $D$, we define
$\prereq{D}=\{ \prereq{\delta} \mid \delta \in D\}$,
$\justif{D}=\{ \justif{\delta} \mid \delta \in D\}$, and
$\conseq{D}=\{ \conseq{\delta} \mid \delta \in D\}$.

A \emph{default theory} is a pair $(D,W)$ where $D$ is a set of 
default rules and $W$ a set of formulas.
A set of conclusions (sanctioned by a given set of default rules and by 
means of classical logic) is called an \emph{extension} of an initial set 
of facts.
More formally, extensions are defined as follows:
\begin{definition}[\cite{Reiter80}]\label{def:dl:extension}
Let $(D,W)$ be a default theory and let $E$ be a set of formulas.
Define
\(
E_1 = W
\)
and, for $n \geq 1$,
\begin{eqnarray*}
E_{n+1} &=& \theory{E_n}
            \;\cup\;
            \left\{\;\gamma\;\left|\;\mbox{$\dr{\alpha}{\beta}{\gamma}$}\in D,
                          \alpha\in E_n,\neg\beta\not\in E
            \right.\right\}.
\end{eqnarray*}
Then, $E$ is an 
{extension} of $(D,W)$ iff $E = \bigcup_{n\in\omega} E_n$.
\end{definition}

\section{Signed Systems}\label{sec:signed}

The basic idea of signed systems is to transform an inconsistent theory into a 
consistent one by renaming propositional variables and then to extend the
resulting signed theory by equivalences using {default logic}.

\subsection{Basic Approach}

Starting with a possibly inconsistent finite theory 
$W\subseteq\lang_\Sigma$, we consider the  default theory 
obtained from
\(
W^\pm
\)
and a set of default rules
\(
D_\Sigma=\{\delta_p\mid p\in\Sigma\}
\)
defined in the following way.
For each propositional letter $p$ in $\Sigma$, we define
\begin{equation}\label{def:default:para}
\delta_p = \paradefneu{p}
\ .
\end{equation}
Using this definition, we define the first family of 
paraconsistent consequence relations:
\begin{definition}\label{def:consequence:unsigned:i}
  Let $W$ be a finite set of formulas in $\lang_\Sigma$ and let $\varphi$ be a
  formula in $\lang_\Sigma$.
  Let $\mathcal{E}$ be the set of all extensions of \
  \(
  (D_\Sigma,W^\pm)
  \).
  For each set of formulas $S\subseteq\lang_{\Sigma\cup\Sigma^\pm}$, 
  let
  \[
  \Pi_S=\{\conseq{\delta_p}\mid p\in\Sigma,
                                \neg\justif{\delta_p}\not\in S\}.
  \]
  Then, we define
  \begin{itemize}
        \item[]
          \(
          W\vdash_c\varphi
          \)
          \ iff\/\
          \(
          \varphi\in\bigcup_{E\in\mathcal{E}}\theory{W^\pm\cup\Pi_E}
          \)
          \hfill {\rm (}credulous unsigned\footnote{The term ``unsigned'' indicates
          that only unsigned formulas are taken into account.}
        consequence{\rm )}
    \item []
          \(
          W\vdash_s\varphi
          \)
          \ iff\/\
          \(
          \varphi\in\bigcap_{E\in\mathcal{E}}\theory{W^\pm\cup\Pi_E}
          \)
          \hfill {\rm (}skeptical unsigned consequence{\rm )}
    \item []
          \(
          W\vdash_p\varphi
          \)
          \ iff\/\
          \(
          \varphi\in\theory{W^\pm\cup\bigcap_{E\in\mathcal{E}}\Pi_E}
          \)
          \hfill {\rm (}prudent unsigned consequence{\rm )}
  \end{itemize}
\end{definition}
For illustration, consider the inconsistent theory
\begin{equation}
  \label{ex:1}
  W = \{ p, q, \neg p \vee \neg q \}.  
\end{equation}
For obtaining the above paraconsistent consequence relations, $W$ is 
turned into the default theory\footnote{For simplicity, we omitted all
  $\delta_x$ for $x\in\Sigma\setminus\{p,q\}$.}
\[
(D_\Sigma,W^\pm)=\big(\{\delta_p, \delta_q\}, \{p^+, q^+, p^- \vee q^-\}\big).
\]
We obtain two extensions, viz.\ $\theory{W^\pm\cup\{\conseq{\delta_p}\}}$ and 
$\theory{W^\pm\cup\{\conseq{\delta_q}\}}$.
The following relations show how the different consequence relations behave: 
$$
W\vdash_c p,\quad W \not\vdash_s p,\quad W \not\vdash_p p,
$$
but, for instance,
$$
W\vdash_c p\vee q, \quad W \vdash_s p\vee q, \quad W \not\vdash_p p\vee q.
$$

For a complement, the following ``signed'' counterparts are defined.
\begin{definition}\label{def:consequence:signed:i}
  Given the prerequisites of
  Definition~\ref{def:consequence:unsigned:i},
  we define
  \begin{itemize}
  \item []
    \(
    W\vdash_c^\pm\varphi
    \)
    \ iff\/\
    \(
    \varphi^\pm\in\bigcup_{E\in\mathcal{E}}\theory{W^\pm\cup\Pi_E}
    \)
    \hfill {\rm (}credulous signed consequence{\rm )}    
  \item []
    \(
    W\vdash_s^\pm\varphi
    \)
    \ iff\/\
    \(
    \varphi^\pm\in\bigcap_{E\in\mathcal{E}}\theory{W^\pm\cup\Pi_E}
    \)
    \hfill {\rm (}skeptical signed consequence{\rm )}
  \item []
    \(
    W\vdash_p^\pm\varphi
    \)
    \ iff\/\
    \(
    \varphi^\pm\in\theory{W^\pm\cup\bigcap_{E\in\mathcal{E}}\Pi_E}
    \)
    \hfill {\rm (}prudent signed consequence{\rm )}
  \end{itemize}
\end{definition}

\subsection{Formal Properties}

As shown in~\cite{Besnard98},
these relations compare to each other in the following way.
\begin{theorem}\label{thm:C:props:i}
Let $C_i$ be the operator corresponding to
\(
C_i(W)=\{\varphi\mid W\vdash_i\varphi\}
\) 
where $i$ ranges over
\(
\{p,s,c\},
\)
and similarly for $C^\pm_i$.
Then, we have
\begin{enumerate}
  \item \(
        C_i(W)\subseteq C^\pm_i(W)
        \);
  \item \(
        C_p(W)\subseteq C_s(W)\subseteq C_c(W)
        \)
        \ and\ 
        \(
        C^\pm_p(W)\subseteq C^\pm_s(W)\subseteq C^\pm_c(W)
        \)  .
\end{enumerate}
\end{theorem}
That is, signed derivability gives more conclusions than unsigned
derivability and within each series of consequence relations the strength of
the relation is increasing. 

Moreover, they enjoy the following logical properties:
\begin{theorem}\label{thm:C:props:ii}
Let $C_i$ be the operator corresponding to
\(
C_i(W)=\{\varphi\mid W\vdash_i\varphi\}
\) 
where $i$ ranges over
\(
\{p,s,c\},
\)
and similarly for $C^\pm_i$.
Then, we have
\begin{enumerate}
\setcounter{enumi}{2}
  \item \(
        W\subseteq C^\pm_i(W)
        \);
  \item \(
        C_p (W)=\theory{C_p (W)}
        \)
        \ and\/\ 
        \(
        C_s (W)=\theory{C_s (W)}
        \);
  \item \(
        C^\pm_i (W)=C^\pm_i(C^\pm_i (W))
        \);
  \item \(
        \theory{W}\neq\lang_\Sigma
        \)
        only if
        \(
        \theory{W}=C_i(W)=C^\pm_i(W)
        \);
  \item \(
        C_i(W)    \neq\lang_\Sigma
        \)
        \ and \ 
        \(
        C^\pm_i(W)\neq\lang_\Sigma
        \);
  \item \(
        W\subseteq W'
        \)
        does not imply
        \(
        C_i(W)\subseteq C_i(W'),
        \)
        \ and\ 
        \(
        W\subseteq W'
        \)
        does not imply
        \(
        C^\pm_i(W)\subseteq C^\pm_i(W')
        \).
\end{enumerate}
\end{theorem}
The last item simply says that all of our consequence relations are
nonmonotonic.
For instance,
we have
\(
C_i(\{A,A\to B\})=C^\pm_i(\{A,A\to B\})=\theory{\{A,B\}},
\)
while neither
\(
C_i(\{A,\neg A,A\to B\})
\)
nor
\(
C^\pm_i(\{A,\neg A,A\to B\})
\)
contains $B$. 

\subsection{Refinements}

The previous relations embody a somewhat global approach in restoring
semantic links between positive and negative literals.
In fact, the application of a rule $\delta_p$ re-establishes the semantic
link between all occurrences of proposition $p$ and its negation
$\neg p$ at once.
A more fine-grained approach is to establish the connections between
complementary occurrences of an atom individually.

Formally, for a given $W$ and an index set $I$\/ assigning different 
indices to all
occurrences of all atoms 
in $W$, define 
\begin{equation}
  \label{def:default:para:connection}
  \delta_p^{i,j}
  =
  \paradefneuC{p}{i}{j}
\end{equation}
for all $p\in\Sigma$ and all $i,j\in I$, provided that $i$ and $j$ refer to 
complementary occurrences of $p$ in $W$, otherwise set
\(
\delta_p^{i,j} = \delta_p
\).
Denote by $D_\Sigma^1$ this set of defaults and by $W^\pm_I$ the result of
replacing each $p^+\in W^\pm$ (resp., $p^-\in W^\pm$) by $p^+_i$ (resp.,
$p^-_i$) where $i$ is the index assigned to the corresponding occurrence,
provided that there are complementary occurrences of $p$ in $W$. 

Finally, abandoning the restoration of semantical links and foremost
restoring original (unsigned) literals leads to the most adventurous approach
to signed inferences.
Consider the following set of defaults, defined for all $p\in\Sigma$ and 
$i,j\in I$,
\begin{equation}
  \label{def:default:para:propositions}
  \delta_p^{i+}
  =
  \paradefneuPp{p}{i}
  \qquad
  \delta_p^{j-}
  =
  \paradefneuPm{p}{j}
\end{equation}
for all positive and negative occurrences of $p$, respectively.
As above,
we use these defaults provided that there are 
complementary occurrences of $p$ in $W$, otherwise use 
\(
\delta_p
\).
A set of defaults of form~(\ref{def:default:para:propositions}) with respect to $W$ is
denoted by $D_\Sigma^2$.

Thus, further consequence relations are defined when $(D_\Sigma,W^\pm)$ in
Definition~\ref{def:consequence:unsigned:i}
is replaced by $(D_\Sigma^1,W^\pm_I)$ or by $(D_\Sigma^2,W^\pm_I)$.
Similar results to Theorem~\ref{thm:C:props:i} and~\ref{thm:C:props:ii}
can be shown for these families of consequence relations.

In the following, we identify all introduced default theories as follows.
Given a finite set $W\subseteq\lang_\Sigma$, the class $\mathcal{D}(W)$ contains 
$(D_\Sigma,W)$, as well as $(D_\Sigma^1,W^\pm_I)$ and $(D_\Sigma^2,W^\pm_I)$  
for any index set $I$.
Furthermore,  $\mathcal{D}=\bigcup_{W\subseteq\lang_\Sigma} \mathcal{D}(W)$ denotes
the class of all possible default theories under consideration. 

\subsection{Hierarchic Extensions}

Whenever a problem instance may give rise to several solutions,
it is useful to provide a preference criterion for selecting a subset of
preferred solutions.
This is accomplished in \cite{Besnard98} by means of a \emph{ranking function}
\(
\varrho: \Sigma\to I\!\!N
\)
on the alphabet $\Sigma$ for inducing a hierarchy on the
default rules in $D_\Sigma$\/:
\begin{definition}\label{def:hierarchy}
Let
\(
\varrho: \Sigma\to I\!\!N
\)
be some ranking function on alphabet $\Sigma$, 
and $(D,V)\in\mathcal{D}$.
We define the {hierarchy} of $D$ with respect to
\(
\varrho
\)
as the partition
\(
\langle D_n \rangle_{n\in\omega}
\)
of $D$  such that
for each
$\delta\in D$ with $\delta$ of form
$\delta_p,\delta^{i,j}_p,\delta^{i+}_p, \delta^{i-}_p$, for
$p\in\Sigma$ and $i,j\in I$, 
$\delta\in D_n$ iff $\varrho(p)=n$ holds.
\end{definition}
Strictly speaking,
\(
\langle D_n \rangle_{n\in\omega}
\)
is not always a genuine partition, since $D_n$ may be the empty set for some
values of $n$.

Such rankings are used for inducing so-called \emph{hierarchic extensions}.
This concept has been introduced to deal with a given partition on the defaults $D$ 
of a default theory $(D,V)$. 

\begin{definition}\label{def:dl:extension:hierarchic}
Let $W$ be a finite set of formulas in $\lang_\Sigma$,
$(D,V)\in\mathcal{D}(W)$,
and $E$ a set of formulas.
Let
\(
\langle D_n \rangle_{n\in\omega}
\)
be the hierarchy of $D$ with respect to some ranking function $\varrho$.

Then,
\(
E = \bigcup_{n\in\omega} E_n
\)
is a hierarchic extension
of\/ $(D,V)$ relative to $\varrho$
if\/
\(
E_1 = V 
\)
and
\(
E_{n+1} \mbox{ is an extension of } (D_{n},E_n)
\)
for all $n\geq 1$.
\end{definition}

Let
\(
\langle D_n \rangle_{n\in\omega}
\)
be the hierarchy of $D$ with respect to some ranking function $\varrho$, 
and let $\mathcal{E}$ be the set of all hierarchic extensions of 
a default theory $(D,V)\in\mathcal{D}$
in Definition~\ref{def:consequence:unsigned:i}.
Then, we immediately get corresponding consequence 
relations $\vdash_{ch}$, $\vdash_{sh}$, and $\vdash_{ph}$. 
Furthermore, applying hierarchic extensions on default theories
$(D_\Sigma,W^\pm)$ in accordance to Definition~\ref{def:consequence:signed:i} yields
new relations $\vdash^\pm_{ch}$, $\vdash^\pm_{sh}$, and $\vdash^\pm_{ph}$.

\bigskip
In concluding this section, let us briefly recapitulate all paraconsistent 
consequence relations introduced so far.
As a basic classification, we have credulous, skeptical and prudent consequence.
For each 
of these relations, we
defined 
unsigned operators, which are invokable on three
different classes  
of default theories 
(viz.\ on 
$(D_\Sigma,W^\pm)$, $(D^1_\Sigma,W^\pm_I)$, and $(D^2_\Sigma,W^\pm_I)$),
either on ordinary extensions ($\vdash_i$) or on hierarchic
extensions ($\vdash_{ih}$), and,
on the other hand, 
signed operators
also relying on ordinary extensions ($\vdash^\pm_i$)
or hierarchic extensions ($\vdash^\pm_{ih}$)
of the default theory
 $(D_\Sigma,W^\pm)$.
This gives in total 18 unsigned and 6 signed paraconsistent 
consequence relations, which shall all be considered in the following two 
sections. 

\section{Reductions} \label{sec:red}

In this section, we show how the above introduced consequence relations can be
mapped into quantified Boolean formulas in polynomial time.

Recall the set $\mathcal{D}(W)$ for finite $W\subseteq\lang_\Sigma$.  
In what follows, we use finite default theories
$$
\mathcal{D}^*(W) = \{ ( D_W, V) \mid (D,V)\in \mathcal{D}(W) \} 
$$
where $D_W  =
  \{ \delta\in D\mid \var{\delta}\cap\var{W}\not =\emptyset\}$.
Hence, $D_W$ contains each default from $D$ having an unsigned atom which 
also occurs in $W$. 

The next subsection shows the adequacy of these default theories.
Afterwards, Section~\ref{sec:qbfs} gives QBF-reductions based
on the finite default theories $\mathcal{D^*}(W)$.

\subsection{Finitary Characterisations}

\begin{lemma}\label{thm:finite}
  Let $W\subseteq\lang_\Sigma$ be a finite set of formulas and
  $(D,V)\in\mathcal{D}(W)$ a default theory.
  Moreover, let 
  $C\subseteq{D}$
  and 
  $C_W=\{ \delta\in C\mid \var{\delta}\cap\var{W}\not =\emptyset\}$.
  Then, 
  \begin{enumerate} 
  \item $\theory{V\cup\conseq{C_W}}\cap\lang_\Sigma=
        \theory{V\cup\conseq{C}}\cap\lang_\Sigma$; and
  \item 
        for each $\varphi^\pm\in\lang_{\Sigma^\pm}$,
        $\varphi^\pm\in\theory{V\cup\conseq{C}}\mbox{\ iff\ }
        \varphi^\pm\in\theory{V\cup\conseq{C_W}\cup\conseq{D_\varphi}}$
where   $D_\varphi=
\{\delta_p\mid p\in \var{\varphi}\setminus\var{W} \}$.
  \end{enumerate} 
\end{lemma}

Both results show that having computed a (possibly hierarchic) extension, one has 
a finite set of generating defaults sufficient for deciding whether a
paraconsistent consequence relation holds. The following result shows
that these sets are also sufficient to compute the underlying
extensions themselves.

\begin{theorem}\label{thm:finite:final} 
Let 
$W$, $(D,V)$, $C$, and $C_W$ be as in Lemma~\ref{thm:finite}, and
let $D_W  =
  \{ \delta\in D\mid \var{\delta}\cap\var{W}\not =\emptyset\}$.

Then,
there is a one-to-one correspondence between the extensions of $(D,V)$ 
and the extensions of $(D_W,V)$.
In particular, 
$\theory{V\cup \conseq{C}}$ is an extension of $(D,V)$
iff
$\theory{V\cup \conseq{C_W}}$ is an extension of $(D_W,V)$.
Similar relations hold for hierarchic extensions as well.
\end{theorem}

The next result gives a uniform characterisation for all default theories
under consideration.
It follows from the fact that, for each $\delta_p$, the consequent
\(
(p\equiv p^+)\wedge(\neg p\equiv p^-)
\)
is actually equivalent to
\(
(p^+\equiv\neg p^-)\wedge(p\equiv p^+)
\),\
and, furthermore, that defaults of form~(\ref{def:default:para:connection})
and (\ref{def:default:para:propositions}) share the property
that their justifications and consequents are identical.
Hence, given $W$ and $I$ as usual,
it holds that $\conseq{\delta}\models\justif{\delta}$, 
for each $\delta\in D$, with $(D,V)\in\mathcal{D}^*(W)$.
\begin{theorem}\label{thm:signed:1}
  Let $W\subseteq\lang_\Sigma$ be a finite set of formulas,
  let $(D,V)\in\mathcal{D}^*(W)$ be a default theory, and
  let $C\subseteq D$.

  Then,
  $\theory{V\cup\conseq{C} }$ is an extension of $(D,V)$
  iff
  $j(C)$ is a maximal subset of $j(D)$ consistent with $V$.
\end{theorem}

Note that the subsequent QBF reductions, obtained on the basis of the above result, 
represent a more compact axiomatics than the encodings given in \cite{Egly00c} 
for arbitrary default theories.

We derive an analogous characterisation for hierarchic extensions.
In fact, each hierarchic extension is also an extension (but not vice versa)
\cite{Besnard98}.
Thus, we can characterise hierarchic extensions of a default theory $(D,V)$
as ordinary extensions, viz.\ by $\theory{W\cup\conseq{C}}$ with $C\subseteq
D$ suitably chosen.
The following result generalises Theorem~\ref{thm:signed:1}
with respect to a given partition on the defaults. In particular, if
$\langle D_n \rangle_{n \in \omega}=\langle D \rangle$,
Theorem~\ref{thm:signed:2} corresponds to Theorem~\ref{thm:signed:1}.

\begin{theorem}\label{thm:signed:2}
  Let $W$, $(D,V)$, and $C$ be given as in Theorem~\ref{thm:signed:1}.

  Then,
  $\theory{V\cup\conseq{C}}$
  is a hierarchic extension of $(D,V)$
  with respect to partition $\langle D_n \rangle_{n \in \omega}$ on $D$
  iff
  for each $i\in \omega$,
  $\justif{D_i\cap C}$
  is a maximal subset of
  $\justif{D_i}$ consistent with
  $V\cup\bigcup_{j<i} \conseq{D_j\cap C}$. 
\end{theorem}

Finally, in order to relate extensions of default theories to paraconsistent 
consequence operators, 
we note the following straightforward observations.
 
Let  $\Pi_S$ be as in Definition~\ref{def:consequence:unsigned:i}.
Then, for each extension $E$ of $(D,V)\in\mathcal{D}(W)$,
there exists a $C\subseteq{D}$ such that $\conseq{C}=\Pi_E$.
However, since we have to check whether a given formula 
is contained in some $\theory{V\cup\Pi_E}$, by Lemma~\ref{thm:finite}
it is obviously sufficient to consider just 
the generating defaults of an extension of the corresponding 
restricted default theory
from $\mathcal{D^*}(W)$.
In view of Theorems~\ref{thm:signed:1} and~\ref{thm:signed:2},
this
immediately implies 
that all paraconsistent consequence relations
introduced so far can be characterised by maximal subsets
of the consequences $\conseq{D}$
of
the corresponding default theory $(D,V)\in\mathcal{D^*(W)}$.
More specifically, credulous and skeptical paraconsistent consequence reduces
to  checking whether a given formula is contained in at least one or
respectively
all such maximal subsets. 
Additionally, prudent consequence enjoys the following property.

\begin{lemma}\label{thm:signed:3}
  Let $W\subseteq\lang_\Sigma$ be a finite set of formulas,
  and $(D,V)\in\mathcal{D}^*(W)$.
        
Then,   for each $\varphi\in\lang_\Sigma$, 
  we have that $W\not\vdash_p\varphi$ 
  {\rm (}resp., $W\not\vdash_{ph} \varphi${\rm )}
  iff 
  there exists a set $C\subseteq D$ such that
  $\varphi\notin\theory{V\cup\conseq{C}}$ and, for each 
  $\delta\in D\setminus C$, there is some extension {\rm (}resp., 
  hierarchic extension{\rm )}
  $E$ of $(D,V)$ such that 
  $\conseq{\delta}\notin E$. 
  An analogous result holds for relations $\vdash^\pm_p$ and
   $\vdash^\pm_{ph}$. 
\end{lemma}

\subsection{Main Construction}\label{sec:qbfs}

We start with some basic QBF-modules. 
To this end, recall the schema $\Fcons{\cdot,\cdot}$ from 
Proposition~\ref{prop:sat-check}. 

\begin{definition}\label{def:all}
  Let $W\subseteq\lang_\Sigma$ be a finite set of formulas and
  $\varphi\in\lang_\Sigma$.
  For each finite default theory $T=(D,V)\in\mathcal{D}^*(W)$, 
  let  
  $D=\{\delta_1\commadots\delta_n\}$,
  and define
  \begin{eqnarray*}
    \Ext{T}          & = & \Fcons{V,\justif{D}} \wedge \bigwedge_{i = 1}^n 
                           \Big(
                           \neg g_i \to \neg\, 
                           \Fcons{V\cup\{\justif{\delta_i}\},\justif{D\setminus\{\delta_i\}}}
                           \Big);
                           \\
    \Cons{T,\varphi} & = & \forall P 
                           \Big( 
                           V \wedge  (G \leq \conseq{D}) \to \varphi
                           \Big),
\end{eqnarray*}
where 
$P$ denotes the set of atoms occurring in $T$ or $\varphi$, and
$G=\{g_i \mid \delta_i \in D\}$ is an indexed set of globally 
new variables corresponding to $D$.
\end{definition}

\begin{lemma}\label{lemma:tests}
  Let $W$, $T=(D,V)$, and $G$ be as in Definition~\ref{def:all}.
  Furthermore,
  for any set $C\subseteq D$,
  define the interpretation $M_{C}\subseteq G$ 
  such that $g_i\in M_{C}$ iff $\delta_i\in C$, for $1\leq i \leq n$.

  Then, the following relations hold:
  \begin{enumerate}
  \item $\theory{V\cup \conseq{C}}$ is an extension of $T$ iff $\Ext{T}$
        is true under $M_{C}$; and
  \item $\varphi\in\theory{V\cup \conseq{C}}$ iff $\Cons{T,\varphi}$
        is true under $M_{C}$,
    for any formula $\varphi$ in $\lang_\Sigma$.
  \end{enumerate}
\end{lemma}

Observe that
the correctness of Condition~1 follows directly from  
Theorem~\ref{thm:trans:1}, since we have that
$\Ext{T}$ is true under $M_{C}$ iff
$j(C)$ is a maximal subset of $j(D)$ consistent with $V$, and, in view of
Theorem~\ref{thm:signed:1}, the latter holds iff $\theory{V\cup\conseq{C}}$
is an extension of $T$.
Moreover, Condition~2 is actually reducible to Proposition~\ref{prop:sat-check}.
Combining these two QBF-modules, we obtain encodings for the basic inference
tasks as follows:

\begin{theorem}\label{thm:alltogether}
  Let $W\subseteq\lang_\Sigma$ be a finite set of formulas,
  $T=(D,V)$ a default theory from $\mathcal{D}^*(W)$ 
        with $D=\{\delta_1\commadots\delta_n\}$, 
  $\varphi$ a formula in $\lang_\Sigma$,
  and 
  $G= \{g_1\commadots g_n\}$
  the indexed set of variables occurring in 
  $\Ext{T}$ and $\Cons{T,\varphi}$. 

  Then,  
  paraconsistent credulous and skeptical consequence relations 
  can be axiomatised by means of QBFs as
  follows:
  \begin{enumerate} 
  \item $W\vdash_c \varphi$ 
    iff\/
    \(
    \models \exists\,G(\Ext{T}\wedge\Cons{T,\varphi})\);
    and
  \item $W\vdash_s \varphi$ 
    iff\/ 
    \(
    \models\neg\exists\,G(\Ext{T}\wedge\neg\,\Cons{T,\varphi})
    \).
  \end{enumerate}

        Moreover, for prudent consequence, let
        $G'=\{g'_i \mid g_i \in G\}$ be an additional
        set of globally new variables and
        $$
        \Psi = 
                   \bigwedge_{i=1}^{n}\Big(\neg g'_i \to
                   \exists\,G\big(\Ext{T}\wedge\neg\,\Cons{T,\conseq{\delta_i}}\big)\Big).
        $$
 Then,
        \begin{enumerate}
        \item[3.] 
  $W\vdash_p \varphi$ 
    iff\/ $\models\neg \exists G' ( \neg \Consp{T,\varphi} \wedge \Psi )$,
   \end{enumerate}
  where $\Consp{T,\varphi}$ 
        denotes the QBF  obtained from 
  $\Cons{T,\varphi}$ 
        by replacing each occurrence of an atom 
        $g\in G$ in $\Cons{T,\varphi}$ by $g'$.
\end{theorem}

In what follows, we discuss the remaining consequence relations under consideration.
We start with signed consequence.
Here, we just have to adopt the calls to 
$\Cons{(D_\Sigma,W^\pm),\varphi}$ 
with respect to 
Lemma~\ref{thm:signed:3},
by adding those 
defaults $\delta_p$ to $W^\pm$ such that 
$p\in\var{\varphi}\setminus\var{W}$.
Observe that in the following theorem this addition is \emph{not}
necessary for the module $\Psi$.
Furthermore, recall that signed consequence is applied only to default theories
$(D_\Sigma,W^\pm)$.

\begin{theorem}\label{thm:alltogether:2}
  Let $W\subseteq\lang_\Sigma$ be a finite set of formulas and
  $\varphi$ a formula in $\lang_\Sigma$.
  Moreover, let 
        $D_W=\{\delta_p\mid p\in\var{W}\}$ and
        $D_\varphi=\{\delta_p\mid p\in\var{\varphi}\setminus\var{W}\}$, 
        with the corresponding default theories
        $T=(D_W,W^\pm)$ and $T'=(D_W,W^\pm\cup\conseq{D_\varphi})$, and
  let $G$, $G'$,  and  $\Psi$ be as in Theorem~\ref{thm:alltogether}.

  Then,  
  paraconsistent signed 
consequence relations 
  can be axiomatised by means of QBFs as
  follows:
  \begin{enumerate} 
  \item $W\vdash^\pm_c \varphi$ 
    iff\/
    \(
    \models\exists\,G(\Ext{T}\wedge\Cons{T',\varphi^\pm})
    \);
  \item $W\vdash^\pm_s \varphi$ 
    iff\/
    \(
    \models\neg \exists\,G(\Ext{T}\wedge\neg\,\Cons{T',\varphi^\pm})
    \); and 
   \item $W\vdash^\pm_p \varphi$ 
    iff\/ $\models\neg \exists G' ( \Psi \wedge \neg \Consp{T',\varphi^\pm}  )$,
   \end{enumerate}
  where, as above,  $\Consp{\cdot,\cdot}$ 
  replaces each $g$ by $g'$.
\end{theorem}

It remains to consider the consequence relations based on hierarchical extensions.
To this end, we exploit the characterisation of Theorem~\ref{thm:signed:2}.
        
\begin{definition}\label{def:hier}
  Let $W\subseteq\lang_\Sigma$ be a finite set of formulas,
  $T=(D,V)$ a default theory from $\mathcal{D}^*(W)$ 
        with $D=\{\delta_1\commadots\delta_n\}$, and
  $P=\langle D_n\rangle_{n\in\omega}$ a partition on $D$. We define
$$
    \ExtH{T,P}        =  \bigwedge_{i\in\omega}
                           \Big( 
                           \Ext{ ( V\wedge\bigwedge_{\delta_j\in D_1\cup\ldots\cup D_{i-1}}
                             ( g_j \to \conseq{\delta_j}) \,,\,\, D_i)}
                           \Big),
$$
where
$G=\{g_i \mid \delta_i \in D\}$ is the same indexed set of globally 
new variables corresponding to $D$ as above appearing in each $\Ext{\cdot}$.
\end{definition}

\begin{lemma}
  Let $W$, $(D,V)$, $G$, and $P$ be as in Definition~\ref{def:hier}.
  Furthermore,
  for any set $C\subseteq D$,
  define the interpretation $M_{C}\subseteq G$ 
  such that $g_i\in M_{C}$ iff $\delta_i\in C$, for $1\leq i \leq n$.

        Then,
   $\theory{V\cup
\conseq{C}}$ is a hierarchic extension of $T$ with respect to $P$ 
iff $\ExtH{T,P}$ is true under $M_{C}$.
\end{lemma}

\begin{theorem}\label{thm:last}
Paraconsistent consequence relations 
$\vdash_{ch}$,
$\vdash^\pm_{ch}$,
$\vdash_{sh}$,
$\vdash^\pm_{sh}$,
$\vdash_{ph}$, and
$\vdash^\pm_{ph}$ are expressible in the same manner as in Theorems~%
\ref{thm:alltogether} and~
\ref{thm:alltogether:2}  by replacing
$\Ext{T}$ with $\ExtH{T,P}$.
\end{theorem}

This concludes the reductions to QBFs.  
Observe that all these reductions are solely built from simple QBF-modules 
like $\Ext{\cdot}$ and $\Cons{\cdot,\cdot}$ and are constructible in polynomial time. 

\section{Complexity Issues}
\label{sec:complexity}

In what follows, we assume the reader familiar with the basic
concepts of complexity theory (cf.\ \egc \cite{Papadimitriou94} for a
comprehensive textbook on this subject).
Relevant for our purposes are the complexity classes $\SigmaP{2}$ and $\PiP{2}$.
$\SigmaP{2}$ is the class of all problems solvable on a
nondeterministic Turing machine in polynomial time having
access to an oracle for problems in $\NP$ 
(the class $\NP$ consists of all decision problems
which can be solved with a nondeterministic Turing machine
working in polynomial time), and
$\PiP{2}$
consists of the problems which are complementary to the
problems in $\SigmaP{2}$, \iec $\PiP{2}=\co\SigmaP{2}$.
Recall that both classes are part of the polynomial
hierarchy. 

In the sequel, we derive complexity results for deciding paraconsistent
consequence in all variants discussed previously.
We show that all considered tasks are located at the second level of the 
polynomial hierarchy. 
This is 
in some sense not surprising, 
because the current
approach relies
on deciding whether a given formula is contained in an extension of a 
suitably constructed default
theory. This problem was shown to be $\SigmaP{2}$-complete by 
Gottlob~\cite{Gottlob92}, even if normal default theories are considered. 
However, 
this completeness result is not directly applicable here because of the
specialised default theories in the present setting. 
Furthermore, for dealing with hierarchic extensions, it turns out that 
the complexity remains at the second level of the polynomial hierarchy as well.
This result is interesting, since the definition of hierarchic extensions is 
somewhat more elaborate than standard extensions.
In any case, this observation mirrors in some sense complexity results 
derived for cumulative default logic (cf.~\cite{Gottlob94a}).

In the same way as the satisfiability problem of classical
propositional logic is the ``prototypical'' problem of $\NP$, \iec 
being an $\NP$-complete problem, the satisfiability
problem of QBFs in \emph{prenex form} possessing $k$ quantifier alternations is
the ``prototypical'' problem of the $k$-th level of the polynomial hierarchy.
\begin{proposition}[\cite{Wrathall76}]
\label{prop:qsat}
Given a propositional formula $\phi$ whose atoms are partitioned 
into $i\geq 1$ sets $P_1\commadots P_i$, deciding 
whether $\exists P_1\forall P_2\ldots\quantifier_i P_i\phi$ 
is true is $\SigmaP{i}$-complete, where $\quantifier_i=\exists$ if $i$ is odd and 
$\quantifier_i=\forall$ if $i$ is even, 
Dually, deciding whether 
$\forall P_1\exists P_2\ldots\quantifier'_i P_i\phi$ is true is $\PiP{i}$-complete,
where $\quantifier'_i=\forall$ if $i$ is odd and 
$\quantifier_i=\exists$ if $i$ is even.
\end{proposition} 
 
Given the above characterisations, we can estimate upper complexity
bounds for the reasoning problems
discussed in Section~\ref{sec:signed}
simply by inspecting the quantifier order of the respective QBF encodings.
This can be argued as follows.
First of all, by applying quantifier transformation rules similar to 
ones in first-order logic, 
each of the above QBF encodings
can be
transformed in polynomial time into a 
QBF in prenex form having exactly one quantifier alternation.
Then, by invoking Proposition~\ref{prop:qsat} and observing
that completeness of a decision problem $D$ for a complexity
class $C$ implies membership of $D$ in $C$,
the quantifier order of the resultant QBFs determines in
which class of the polynomial hierarchy the corresponding
reasoning task belongs to.

Applying this method to our considered tasks,
we obtain that credulous paraconsistent
reasoning lies in $\SigmaP{2}$, 
whilst
skeptical and prudent paraconsistent reasoning
are in $\PiP{2}$.
Furthermore, note that the
QBFs expressing paraconsistent reasoning using the
concept of hierarchical extensions share exactly the same quantifier 
structures as those using ordinary extensions.

\begin{table}[t]
\caption{Complexity results for all paraconsistent consequence relations.}
\label{table:compl}
\begin{center}
\begin{tabular}{|l|c|c|c|}
\hline
&  $T_0=(D_\Sigma,W^\pm)$ & $T_1=(D_\Sigma^1,W^\pm_I)$ &
$T_2=(D_\Sigma^2,W^\pm_I)$ \\
\hline
$\vdash_c$ & $\SigmaP{2}$ & $\SigmaP{2}$ & $\SigmaP{2}$ 
\\
$\vdash_s$ & $\PiP{2}$ & $\PiP{2}$ & $\PiP{2}$ 
\\
$\vdash_p$ & $\PiP{2}$ & in $\PiP{2}$ & in $\PiP{2}$ 
\\
$\vdash^\pm_c$ & $\SigmaP{2}$&-&-
\\
$\vdash^\pm_s$ & $\PiP{2}$&-&- 
\\
$\vdash^\pm_p$ & in $\PiP{2}$ &-&- 
\\
$\vdash_{ch}$&
$\SigmaP{2}$ & $\SigmaP{2}$ & $\SigmaP{2}$
\\
$\vdash_{sh}$&
$\PiP{2}$ & $\PiP{2}$ & $\PiP{2}$
\\
$\vdash_{ph}$&
$\PiP{2}$ & in $\PiP{2}$ & in $\PiP{2}$
\\
$\vdash^\pm_{ch}$
& $\SigmaP{2}$
&-&-\\
$\vdash^\pm_{sh}$
& $\PiP{2}$ 
&-&-\\
$\vdash^\pm_{ph}$
& in $\PiP{2}$ 
&-&-\\
\hline
\end{tabular}
\end{center}
\end{table}

Concerning lower complexity bounds, it turns out that most 
of the above given estimations are \emph{strict}, \iec the
considered decision problems are hard  for the respective
complexity classes.
The results are summarised in Table~\ref{table:compl}. 
There, all entries denote completeness results, except where a membership 
relation is explicitly stated.
The following theorem summarises these relations:

\begin{theorem}
The complexity results in Table~\ref{table:compl} hold both for 
ordinary as well as for hierarchical extensions of $T_i$ {\rm (}$i=0,1,2${\rm )} 
as underlying inference principle.
\end{theorem}

Some of these complexity results have already been shown elsewhere.
As pointed out in~\cite{Besnard98}, prudent consequence, $W\vdash_p\varphi$, 
on the basis of the default theory $(D_\Sigma,W^\pm)$ captures the 
notion of \emph{free-consequences} as introduced in~\cite{Benferhat93}.
This formalism was shown to be $\PiP{2}$-complete in~\cite{Cayrol98}.

Finally, \cite{Marquis02} considers the 
complexity of
a number of different paraconsistent
reasoning principles, among them
the completeness results
for $\vdash_s$ and $\vdash^\pm_s$. 
Moreover, that paper extends the intractability results
to some restricted subclasses as well.

\section{Discussion}\label{sec:discussion}

We have shown how paraconsistent inference problems within the framework of
signed systems can be axiomatised by means of quantified Boolean formulas.
This approach has several benefits:
First, the given axiomatics provides us with further insight about how
paraconsistent reasoning works within the framework of
signed systems.
Second, this axiomatisation allows us to furnish upper bounds
for precise complexity results,
going beyond those presented in \cite{Marquis02}.
Last but not least, we obtain a straightforward implementation technique of
paraconsistent reasoning in signed systems by appeal to existing 
QBF solvers.

For implementing our approach, we rely on the existing system 
\QUIP~\cite{Egly00c,Egly00e}.
The general architecture of \QUIP\ consists of three parts, 
namely the {\tt filter} program,
a QBF-evaluator, and the interpreter {\tt int}.
The input filter translates the given problem description 
(in our case,
a signed system and a specified reasoning task) into
the corresponding quantified Boolean formula, which is then sent to the 
QBF-evaluator. 
The current version of \QUIP\ provides interfaces to most of the 
currently available QBF-solvers. 
The result of the QBF-evaluator is interpreted by {\tt int}. 
Depending on the capabilities of the employed QBF-evaluator, 
{\tt int} provides an explanation in terms of the underlying problem 
instance. 
This task relies on a protocol mapping of internal variables 
of the generated QBF into concepts of the problem description.

\let\proof=\goodoldproof
\let\endproof=\goodoldendproof

\end{document}